\documentclass[12pt,preprint]{aastex}
\def\be{\begin{equation}}
\def\ee{\end{equation}}

\def\msun{M_{\odot}}

\def\be{\begin{equation}}
\def\ee{\end{equation}}
\catcode`\@=11 
\def\@versim#1#2{\vcenter{\offinterlineskip
        \ialign{$\m@th#1\hfil##\hfil$\crcr#2\crcr\sim\crcr } }}
\def\lsim{\mathrel{\mathpalette\@versim<}}
\def\gsim{\mathrel{\mathpalette\@versim>}}

\shorttitle{Hot One-Temperature Accretion Flows}
\shortauthors{Yuan, Taam, Xue, \& Cui}
\begin{document}

\title{Hot One-Temperature Accretion Flows Revisited}

\author{Feng Yuan}
\affil{Department of Physics, Purdue University, West Lafayette, IN 47907;\\
Shanghai Astronomical Observatory, Chinese Academy of Sciences, 
80 Nandan Road, Shanghai 200030, China}
\email{fyuan@shao.ac.cn}

\author{Ronald E. Taam}
\affil{Department of Physics and Astronomy, Dearborn Observatory, 
Northwestern University, 2131 Tech Drive, Evanston, IL 60208}
\email{taam@tonic.astro.northwestern.edu}

\author{Yongquan Xue and Wei Cui}
\affil{Department of Physics, Purdue University, West Lafayette, IN 47907}
\email{xuey,cui@physics.purdue.edu}

\begin{abstract}

The effectiveness of the thermal coupling of ions and electrons in the
context of optically thin, hot accretion flows is investigated in a
phenomenological approach.  In the limit of complete coupling, 
we focus on the one-temperature accretion flows around black holes.  
Based on a global analysis, the results are compared with two-temperature 
accretion flow models and with the observations of black hole sources. 
A comparison of the 
properties of the one-temperature solutions with that of the two-temperature 
solutions reveals many features that are quite similar. That is, hot 
one-temperature solutions are found to exist for mass flow rates less than 
a critical value; i.e., $\dot{M}\la 10\alpha^2\dot{M}_{\rm Edd}$, where 
$\dot{M}_{\rm Edd}= L_{\rm Edd}/c^2$ is the Eddington accretion rate.  
At low mass flow rates, $\dot{M}\la 10^{-3}\alpha^2 \dot{M}_{\rm Edd}$, 
the viscous energy is mainly balanced by the advective {\em cooling}, i.e., 
the solution is in the advection-dominated accretion flow (ADAF) regime. 
On the other hand, at higher rates, $10^{-3}\alpha^2 \dot{M}_{\rm Edd} 
\la \dot{M} \la 10\alpha^2\dot{M}_{\rm Edd}$, radiative cooling is effective 
and is mainly balanced by advective {\em heating}, placing the solution 
in the regime of luminous hot accretion flow (LHAF). At the highest 
mass flow rates, $\dot{M} \ga 10\alpha^2 \dot{M}_{\rm Edd}$, the accretion 
flow collapses at a transition radius (which increases with increasing 
mass flow rate) with only the standard optically thick and geometrically 
thin disk solution existing in the innermost regions.  To test the 
viability of the one-temperature models, we have fitted the spectra of 
the two black hole sources, Sgr A* and XTE J1118+480, which have  
been examined successfully with two-temperature models.  It is found 
that the one-temperature models do not provide acceptable fits to 
the multi-wavelength spectra of Sgr A* nor to XTE J1118+480 as 
a result of the higher temperatures characteristic of the one-temperature 
models.  It is concluded that the thermal coupling of ions and electrons 
cannot be fully effective and that a two-temperature description is required 
in hot accretion flow solutions.
\end{abstract}


\section{Introduction}
 
It has been recognized that accretion disks surrounding black holes can
be described in terms of standard optically thick and geometrically thin
disks (Shakura \& Sunyaev 1973), slim disks (Abramowicz et al. 1988), or
optically thin, geometrically thick disks (Rees et al. 1982). In
contrast to the former descriptions, the latter are hot and can be
described in terms of either an advection-dominated accretion flow
(hereafter ADAF; Ichimaru 1977; Rees et al. 1982; Narayan \& Yi 1994,
1995; Abramowicz et al. 1995; Chen et al. 1995; 
see Narayan, Mahadevan, \& Quataert 1998 for a review) or a
luminous hot accretion flow (LHAF; Yuan 2001, 2003). For a wide 
range of mass accretion rates, both the cool, optically thick and hot, 
optically thin solutions are possible. The particular solution which is 
achieved in reality may well depend on, e.g., initial conditions of the 
accretion flow (Narayan \& Yi 1995). If the temperature 
of the accretion material is virial at the outer boundary, the hot 
accretion solution may be achieved.

The global ADAF solutions occur for accretion rates satisfying $\dot{M}\la 
10\alpha^2 \dot{M}_{\rm Edd}$, where $\dot{M}_{\rm Edd}=L_{\rm Edd}/c^2$ is 
the Eddington accretion rate and $\alpha$ is the viscous parameter. The ion 
energy equation reads $q_{\rm adv}= q_{\rm vis}-q^-$, with $q_{\rm adv}$, 
$q_{\rm vis}$ and $q^-$ corresponding to the rates of 
energy advection, viscous heating and (radiative or Coulomb collision) cooling, 
respectively. In a typical ADAF, $q^- \ll q_{\rm vis}\approx q_{\rm adv}$, 
i.e. the viscous heating is balanced by advection cooling.   
With the increase of accretion rate, $q^-$ increases faster than 
$q_{\rm adv}$. When $\dot{M}\approx 10\alpha^2 \dot{M}_{\rm Edd}$, $q^-$ becomes
sufficiently effective that $q_{\rm vis} \approx q^- \gg q_{\rm adv}$.
This accretion rate is called the critical rate of ADAF (Narayan,
Mahadevan \& Quataert 1998). Previously only standard optically thick and geometrically 
thin disks were thought to be possible above this critical rate. However, 
Yuan (2001; 2003) found that 
a new hot accretion solution 
exists in this regime, i.e, the so-called luminous hot accretion flow (LHAF). 
The equations describing LHAF are completely identical with those of
ADAF, LHAF is therefore a natural extension of ADAF to accretion rates above  
$10\alpha^2 \dot{M}_{\rm Edd}$. In an LHAF, $q_{\rm vis} \la q^-$. 
Then why can the accretion flow still be hot? Note that
$q_{\rm adv}\equiv q_{\rm int}-q_{\rm com}$, where $q_{\rm int}(\equiv
\rho v\frac{d\epsilon_i}{dr}\propto 
v \frac{dT}{dr})$ and $q_{\rm com}$
denote the gradient of internal energy of the gas and the compression work 
respectively.
When $\dot{M}\ga 10\alpha^2 \dot{M}_{\rm Edd}$ (but not too large), we 
have $q_{\rm com}+q_{\rm vis}>q^-$. Therefore $q_{\rm int}>0$,
which means $\frac{dT}{dr}<0$ (note the radial velocity $v<0$). 
This means that the accretion flow will remain hot if it starts out hot,
same as the case of an ADAF. This is the physics of LHAF.
Under the approximation of 
a self-similar ADAF solution,
we roughly have $q_{\rm com}\approx 3 q_{\rm vis}$ (ref. eqs. 8 \& 9 in 
Yuan 2001 and set $\beta=0.9$ in the subsequent 
equation of $T_i$) so the accretion rate
of an LHAF can be $\sim 4$ times higher than the critical rate of ADAF.
In an LHAF, $q_{\rm adv}=q_{\rm vis}-q^-< 0$, 
advection plays a heating rather than a cooling role. In other words, the entropy 
of the accretion flow in an LHAF is converted into radiation, similar to the cases 
of spherical accretion and cooling flow in galactic clusters. We would like to note that
since the advection factor is not constant with radii, the flow can be an ADAF in 
some region while an LHAF in another region, especially when $\dot{M}$ 
is greater than the critical rate of an ADAF. The LHAF solution 
was not found in previous work because it was a priori assumed that the advection 
factor must be positive (e.g., Narayan \& Yi 1995; Abramowicz et al. 1995). 
The ADAF solution has been shown to be successful in explaining 
low-luminosity AGNs or the quiescent state of black hole candidates (BHCs; 
see Narayan 2005 for a recent review), while the LHAF solution shows promise 
in explaining normal AGNs and the hard state of some BHCs (Yuan \& Zdziarski 2004).  

A two-temperature configuration in which the ions are much hotter than
electrons is assumed in the above hot accretion flow models because
Coulomb collisions are not sufficiently effective to thermally couple
these species to force a one temperature plasma. Although the mean free
path of particles for the physical conditions in the hot accretion flow
is often comparable to the size of the accretion flow, collective
plasma effects may alter the description. Whether collective effects
transfer energy from the ions to the electrons on a timescale that is
shorter than the accretion timescale to equilibrate the electrons and ions 
is unknown. If efficient, it would call into question the two-temperature 
assumption (see e.g., Phinney 1981). A possible plasma instability 
mechanism facilitating efficient energy transfer in this regime has been  
suggested by Begelman \& Chiueh (1988). However, it is unclear whether 
this mechanism is sufficiently efficient to eliminate the two-temperature 
nature of the ADAF (Narayan \& Yi 1995). 

In order to determine the effectiveness of the thermal coupling between
ions and electrons, we adopt a phenomenological approach and compare the
different predictions of both one- and two-temperature models with
observations of confirmed black hole sources. It has been thought that 
one-temperature (hereafter one-T) models may not be applicable to observations, 
primarily because the critical mass flow rate below which solutions were found 
to exist was thought to be too low (e.g. $10^{-5}\dot{M}_{\rm Edd}$)
to be of astrophysical interest since the corresponding luminosity is well below 
that of almost all of the astrophysical objects we know (Narayan, Mahadevan 
\& Quataert 1998; Esin et al. 1996). This is correct for the ADAF regime, 
where energy advection represents cooling, however, the critical accretion
rate and corresponding luminosity are several orders of magnitude higher when
the solutions are extended to the LHAF regime where energy advection represents
heating (i.e., the entropy is converted into radiation; see \S 2). 
Another potential undesirable characteristic of such models 
is that the plasma predicted in a one-T model is too hot compared to the electron 
temperature predicted in a two-temperature (hereafter two-T) model with a small 
$\delta$. Here $\delta$ is the of the turbulent viscous energy that directly 
heats the electrons. Considering the theoretical uncertainties of its value (e.g., 
Quataert \& Gruzinov 1999), in the original ADAF model, it was assumed 
to be very small, $\delta \approx 10^{-3}$ or $10^{-2}$.  However, the recent
excellent observational data on the supermassive black hole located at the 
center of our Galaxy, Sgr A*, places strict constraints on the ADAF model and 
the data require $\delta\sim 0.5 \gg 10^{-2}$ (Yuan, Quataert \& Narayan 2003). This
high value of $\delta$ implies that the required electron temperature is 
much higher than predicted in the original ADAF model with a small $\delta$, 
raising the question of whether the accretion flow can be described by a one-T model. 

Previous theoretical studies on one-T accretion flows were either local or did 
not consider the important effects of synchrotron emission and its 
Comptonization on the temperature of the accretion flow.  In spite of these 
limitations, the early pioneering work of Narayan \& Yi (1994), Abramowicz et 
al. (1995), and Chen et al. (1995), focusing on local solutions with simplified
radiative processes, led to important insights on the general nature and 
character of the hot accretion flow solutions and, in particular, on the 
existence of a critical accretion rate and its scaling with $\alpha^2$.  
Subsequently, Esin et al. (1996) took into account 
all the important radiative processes in their study of the dynamics of hot 
one-T accretion flow, but their calculations were based on a local rather than 
a global analysis. Since significant differences in the solutions are expected 
in the innermost regions between calculations based on a local and a global 
analysis, Narayan, Kato \& Honma (1997) and Chen, Abramowicz \& Lasota (1997) 
obtained the global solution of one-T accretion flow, but again, using very 
simplified descriptions for the radiative losses. Most importantly, common 
to all previous work on one-T models, only the ADAF regime was considered, 
neglecting the LHAF regime.  Based on the results of two temperature models 
(see Yuan 2001), the maximum accretion rate of hot one-T accretion flow 
solutions may be much higher. 

Accordingly, in this paper, we report on the results of global one-T accretion 
flow solutions in which all the important nonthermal and thermal radiative
processes are included.  In the next section, we describe the results of our 
calculations and compare them with two-T solutions.  We find that both the 
electron temperature and the highest accretion rate predicted by a one-T 
model and a two-T model with $\delta=0.5$ are quite similar. To examine whether
one-T accretion flow can be realized in nature, the spectra obtained from the 
one-T models are compared with the
observed spectra of the two black hole sources for which extensive observational
data have been obtained (viz., Sgr A* and XTE J1118+480) in \S3.  Finally, we 
summarize our results in \S4.

\section{Dynamics of one-temperature hot accretion flows}

We consider a steady state axisymmetric accretion flow, employing the
height-integrated set of equations for the hot optically thin inner 
regions of an accretion disk. The transition from the cool, optically 
thick outer regions of the disk to the accretion flow studied in this paper
is not fully understood, although processes such as evaporation (Meyer \& 
Meyer-Hofmeister 1994) and turbulent conduction (Manmoto \& Kato 2000) 
have been proposed as important in affecting such a transition. More 
recent investigations by Turolla \& Dullemond (2000) and Spruit \& 
Deufel (2002) show that various forms of evaporation can be effective 
in forming such optically thin inner disks. 

The mass flow rate in the inner hot disk can be expressed in the form 
suggested by Blandford \& Begelman (1999) as \be \dot{M}=-4\pi R H \rho v
= \dot{M}_{\rm out}\left(\frac{R}{R_{\rm out}} \right)^s, \ee where
$\dot{M}_{\rm out}$ is the mass flow rate at the outer boundary of the
flow.  Departures from a constant mass flow rate in the disk associated
with outflow or the onset of convection can be modelled using $s \ne 0$;
however, the dynamics for a constant mass flow rate (i.e., we set $s=0$)
is calculated in the present paper, since we focus on the 
differences of the dynamics between
the one and two-temperature accretion flows. On the other hand, account
of deviations from a constant mass flow rate will be included in fitting
theoretical spectra to the observed spectra.  The other two equations
required for a description of the flow, namely the radial and azimuthal
components of the momentum equation, are respectively \be v
\frac{dv}{dr}= -\Omega_{\rm k}^2 r+\Omega^2 r-\frac{1}{\rho}\frac
{dp}{dr}, \ee \be v(\Omega r^2-j)= \alpha r \frac{p}{\rho}. \ee Here,
$v$ is the radial velocity, $\Omega_k$ is the Keplerian angular
velocity, $\rho$ is the local gas density, $p$ is the total pressure,
and $j$ is an eigenvalue corresponding to the specific angular momentum
loss at the inner boundary.  For ease of comparison with previous
investigations, we adopt the Paczy\'nski \& Wiita (1980) relation for
the gravitational potential to approximate the space time geometry of a
Schwarzschild black hole.

For the two-T accretion flow, the energy equations for electrons and
ions are, \be q_{\rm adv,e}\equiv \rho v \left(\frac{d
\varepsilon_e}{dr}- {p_e \over \rho^2} \frac{d \rho}{dr}\right) =\delta
q^++q_{ie}-q^-, \ee \be q_{\rm adv,i}\equiv \rho v \left(\frac{d
\varepsilon_i}{dr}- {p_i \over \rho^2} \frac{d \rho}{dr} \right)
=(1-\delta)q^+-q_{ie}. \ee 
Here $\varepsilon_e$$(\varepsilon_i)$ is 
the internal energy of electrons (ions) per unit mass of the gas, and 
$p_e(p_i)$ is the pressure due to electrons (ions). They are defined as,

$\varepsilon_e=\frac{1}{\gamma_e-1}\frac{kT_e}{\mu_e m_{\mu}}$
($\varepsilon_i=\frac{1}{\gamma_i-1}\frac{kT_i}{\mu_i m_{\mu}}$), 
$p_e=\frac{\rho}{\mu_e}\frac{k}{m_{\mu}}T_e$
($p_i=\frac{\rho}{\mu_i}\frac{k}{m_{\mu}}T_i$), 

where $m_\mu$ is the mass of the atom.
$q^-$ is the electron cooling rate (including the
synchrotron and bremsstrahlung process and their 
Comptonization; see Yuan 2001 for details),
$q_{ie}$ is the Coulomb energy exchange rate between
electrons and ions (Stepney \& Guilbert 1983), and $q^+$ is the net
turbulent heating rate.  These latter two rates are given by 
\be
q_{ie}=\frac{3}{2}\frac{m_e}{m_p}n_e n_i\sigma_{\rm T}c\frac{kT_i-kT_e}
{K_2(1/\theta_e)K_2(1/\theta_i)}{\rm ln}\Lambda 
\left[\frac{2(\theta_e+\theta_i)^2+1}{\theta_e+\theta_i}
K_1\left(\frac{\theta_e+\theta_i}{\theta_e\theta_i}\right)+
2K_0\left(\frac{\theta_e+\theta_i}{\theta_e\theta_i}\right)\right],
\ee
\be
q^+=-\alpha pr\frac{d\Omega}{dr}.
\ee
Here $T_i$ and $T_e$ are the ion
and electron temperatures, $\theta_i\equiv kT_i/m_ic^2$,
$\theta_e\equiv kT_e/m_ec^2$, $K_0, K_1, K_2$ are modified Bessel 
functions of the second kind of order 0, 1, and 2, respectively;  
${\rm ln}\Lambda$ is the Coulomb logarithm. For a one-T flow,
$T_e=T_i\equiv T$, and the equations are identical to those describing the
two-T flows, except that the energy equation, which is the sum of the
above two equations, takes the form \be q_{\rm adv}\equiv \rho v
\left(\frac{d(\varepsilon_e + \varepsilon_i)}{dr}- {p_i+p_e \over
\rho^2} \frac{d \rho}{dr}\right) = q^+ - q^-. \ee The contribution due
to radiation pressure to the total pressure is not included since it is
generally, at most, $10\%$ of the gas pressure for both the one and
two-T models.  The effect of electron-positron pairs in ADAFs
has only been studied very approximately within the framework of a 
local treatment.  In this case,  the effect of such pairs was found to 
be unimportant (e.g., Bj\"ornsson et al. 1996; Kusunose \& Mineshige 
1996; see Bj\"ornsson 1998 for a review). In the present paper, for 
simplicity we neglect this effect; however, it should be kept in mind 
that the inclusion of pairs in {\em global hot accretion flows (including 
ADAFs and LHAFs)} should be investigated carefully in the future. The system
of equations for the one-T model (eqs. 1, 2, 3, and 8) are solved
subject to the appropriate boundary conditions (see below), to obtain
the characteristics of the one-T hot accretion flow as described by the
spatial variation of the ion and electron temperatures, optical depth,
and inflow velocity.  For our model calculations, we adopt as parameters
$M=10\msun$, $\alpha=0.3$, and $\beta=0.9$.  Here $\beta$ represents the
ratio of the gas pressure to the total pressure (the sum of gas and magnetic 
pressure). The outer boundary is
chosen to lie at $R_{\rm out}=10^3R_s$, where we assume $T_{\rm out}=
10^9$K and $v/c_s=0.5$ (corresponding to an angular velocity
$\Omega_{\rm out}\sim 0.4\Omega_k$). 

The variation of the temperature, $T$, optical depth, $\tau\equiv\sigma_Tn_e H
$, the ratio of the scale height to radius, $H/R$, and the
advection factor, $f_{\rm one}\equiv q_{\rm adv}/q^+$, are illustrated
as a function of radius for a range of mass flow rates in Figs. 1 -- 4.
The solid lines denote the solutions for the one-T model.  For
comparison, the results for the two-T model with $\delta=0.01$
(dot-dashed line) and $\delta=0.5$ (dashed line) are also shown. 
Note in the temperature plot, only the electron temperature $T_e$ is presented.
The outer boundary conditions used for the two-T case are similar to the
one-T case.  The only difference between the two cases is the 
introduction of an additional outer boundary condition on the electron 
temperature, which is $T_e=4\times 10^8$K. We point out that the effect of 
outer boundary condition (OBC) is unimportant for $\delta=0.5$ since 
the viscous heating, which plays an 
important role in determining the electrons temperature, is a local term. 
On the other hand, the OBC is important in determining the electron 
temperature for $\delta=0.01$. In this case, the electron temperature differs 
for different OBCs, but it is always lower than that of the case for 
$\delta=0.5$. Overall, the electron temperature for the two-T model is always 
smaller than that for one-T model.  For comparison, only the
electron temperature $T_e$ is shown in the temperature plots.  
The advection factor for the two-T flow is defined as $f_{\rm two}\equiv
q_{\rm adv,i}/((1-\delta)q^+)$. 

The properties of the accretion flow at a mass flow rate of
$\dot{M}=10^{-5}\dot{M}_{\rm Edd}$ are illustrated in Fig. 1.  It can be
seen that the plasma temperature $T$ is higher than the electron
temperature in the two-T solution.  For example, $T\sim 2 T_e$ for
$\delta=0.5$ and $T\sim 10 T_e$ for $\delta=0.01$ at $R\sim 10R_s$. 
Specifically, at such a low accretion rate, the radiative losses are
ineffective, leading to the temperature $T$ approaching the virial value. 
This leads to similar values of $H/R$ in the solutions since $H \propto
T^{1/2}$ (for the one-T model) and $H \propto T_i^{1/2}$ (for the two-T 
model).  In addition to the difference in electron temperature, the advection
factor is distinctly different between the one and two-T flows.  In the
inner region of the flow, $f_{\rm two}\approx 1$ while $f_{\rm one}<1$. 
This is attributable to the fact that the radiative losses in a one-T
accretion flow are much more effective than in a two-T flow because
of the higher electron temperatures in the flow.  We note, however, that the
solution is still advection-dominated. 

To determine the behavior of the accretion flow of one-T models at
higher mass flow rates, the flow rate was increased to
$\dot{M}=10^{-3}\dot{M}_{\rm Edd}$.  Upon comparison with Fig. 1, the
temperature $T$ is decreased, especially in the region $R< 10R_s$, due to
the increased gas densities.  These lower temperatures lead to an
accretion flow which becomes geometrically thinner in the innermost
regions.  We also note that the advection factor $f_{\rm one}<0$ at the
inner region.  As in the case of two-T LHAF model (Yuan 2001), this
result is due to the fact that the radiation is so strong that the
viscous dissipation rate of energy, alone, can not balance it; rather,
the radiation is balanced mainly by the advection {\em heating}, i.e.,
the entropy of the flow is converted into radiation. Specifically, it
can be seen from eq. (8) that the advection term $q_{\rm adv}$ is
composed of two terms, namely a term proportional to the gradient of the
internal energy of the accretion flow and the compression work,
$q_{\rm adv}=q_{\rm int}-q_{\rm comp}$.  Although $q^+<q^-$, the sum of
$q^+$ and the compression work can be larger than $q^-$. In this case,
the gradient of the internal energy ($d(\epsilon_e+\epsilon_i)/dr$)
is still negative (note that $v$ is negative for accretion), as in the case
of ADAF.  Thus, the flow remains hot. 

Under the assumption that $f_{\rm adv}>0$ throughout the flow, 
Esin et al. (1996) only obtained ADAF-type solutions in contrast to 
the LHAF-type solutions found in the present work. For 
their assumption, the hot accretion solution exists only for $\dot{M}
\la 10^{-2}\alpha^2\dot{M}_{\rm Edd}$, while we find that 
$f_{\rm adv} \ga 0$ for $\dot{M}\la 10^{-4}\dot{M}_{\rm 
Edd}$ and $\alpha=0.3$. The two results are qualitatively consistent, the 
factor of 10 discrepancy arises because Esin et al. (1996) adopted a local 
analysis while a global approach is adopted in the present investigation. 

The profiles for $T$, $\Sigma$, $H/R$, and $f$ are shown in Fig. 3 for a
further increase in the mass flow to $\dot{M}=10^{-1}\dot{M}_{\rm Edd}$
. Compared to Fig. 2, it can be seen that the trends established for
higher mass flow rates are further confirmed.  Namely, the temperature
and scale height of the flow decrease further, and the spatial extent of
the region characterized by a negative $f_{\rm adv}$ becomes larger as a
result of the increasing effectiveness of radiative cooling. 

Upon comparison of Figs. 1, 2, and 3, it is found that the temperature
of one-T model approaches $T_e$ of the two-T models as the mass flow
rate is increased.  The physical reason for this result stems from the
fact that for one-T models (ref. eq. (8)), the two heating terms,
compression work and viscous dissipation, are both proportional to
$\dot{M}$; while for the two-T models (ref.  eq. (4)), the additional
heating term, $q_{\rm ie}$, is proportional to $\dot{M}^2$ (ref. eq. (6)).
Thus, the total heating rate for the two-T models increases
faster with increasing $\dot M$ than the case for one-T models. 
Therefore, $T_e$ decreases more slowly for the two-T models as compared
to the one-T models as the mass flow rate is increased. 

For a further increase in the value of $\dot{M}$, it is found that the
hot one-T solutions do not exist above the critical accretion rate,
$\sim \dot{M}_{\rm Edd}$ or $10\alpha^2\dot{M}_{\rm Edd}$.  The solid
lines (thick and thin) in Fig. 4 show two
possibilities for the one-T models for this critical rate, corresponding
to different outer boundary conditions (OBC). The OBC of the 
thin solid line is as before, i.e., $T=10^9 {\rm K}, v/c_s=0.5$; while that
of the thick solid line is $T=1.2\times 10^9 {\rm K}, v/c_s=0.5$. We note that 
the OBC of the thin solid line corresponds to a larger angular velocity 
($\sim 0.44 \Omega_k$) than the thick solid line ($\sim 0.4 \Omega_k$). 
As a result, the thick solid 
line solution is of Bondi-type while the thin solid line is of disk-type. 
The Bondi-type solutions have larger sonic radii and thus larger radial 
velocity in the inner region of the accretion flow than the disk-type
(ref. the solid line in Fig. 6; see Yuan 1999 or Abramowicz \& Zurek 
1981 for discussions on the Bondi-type and disk-type accretion modes). 
A larger radial velocity results in a smaller surface density, 
according to eq. (1). Therefore, the thick solid line solution can 
extend to the horizon of the black hole while the thin solid line 
solution does not.  In the latter case, the radiative cooling 
of the latter is sufficiently effective because of the large surface 
density that even the sum of compression work and viscous dissipation 
can not balance it. Thus, the hot accretion flow 
denoted by the thin solid line collapses onto the equatorial plane, forming a
standard optically thick and geometrically thin accretion disk (Shakura
\& Sunyaev 1973).  As the mass accretion rate is increased, the radius
at which the transition to a standard thin disk occurs is increased.  This
configuration represents the typical geometry for one-T solutions at
high mass flow rates ($\dot{M}> 10\alpha^2 \dot{M}_{\rm Edd} \sim
\dot{M}_{\rm Edd}$) and is analogous to the case of
two-T LHAF solutions above its critical mass flow rate (Yuan 2001; see
also Begelman, Sikora \& Rees 1987).  

As in Figs. 1-3, the dashed and
dot-dashed lines in Fig. 4 are for two-T solutions with $\delta=0.5$ and
$0.01$, respectively.  It is interesting to note that the difference in
temperatures between the two cases ($\delta=0.01$ and 0.5) is reduced
when $\dot{M}$ is higher because the weight of the viscous heating
in the electron energy equation is smaller and Coulomb collisions
play a more important role.  Examination of the advection plot from Fig.
4 indicates that $\dot{M}= \dot{M}_{\rm Edd}$ is roughly the critical
mass flow rate for two-T ADAF models. In fact, in a restricted region of the
dashed line, the advection factor is negative. Above $\dot{M}_{\rm
Edd}$, the two-T hot solution still exists, becoming an LHAF. The dotted
line in Fig. 4 shows an example of LHAF solution for a mass accretion
rate of $\dot{M}=3\dot{M}_{\rm Edd}$. This value of $\dot{M}$ is near
the maximum rate of an LHAF for a two-T model.  Above this value, the
inner region will collapse (Yuan 2001). Thus, the maximum accretion rate
below which the hot accretion solution exists for a two-T flow is about
3 time higher than that of a one-T flow.

The bolometric luminosities emitted by the models are presented as a
function of the mass flow rate in Fig. 5. As expected, when $\dot{M}$ is small,
the efficiency of the one-T models is much higher than that of two-T models 
with $\delta=0.01$ and higher to a lesser degree than that
of $\delta=0.5$ two-T models, because $T$ in one-T models is significantly 
higher than $T_e$ in two-T models. In this regime, the efficiency
of two-T models increases faster than that of one-T models as the mass
accretion rate is increased. This reflects the fact that, as argued above, 
the Coulomb collision in the electron energy equation increases faster than the
viscous dissipation rate and compression work, which are the heating
sources for one-T models, as $\dot M$ is increased. For sufficiently
high mass accretion rates, $\dot{M}=\dot{M}_{\rm Edd}$, the
efficiency of the Bondi-type one-T model (denoted by the thick solid 
line in Fig. 4) is lower than that of two-T models (denoted by the 
dashed and dot-dashed lines in Fig. 4). This is due to the 
fact that although $T$ is higher than $T_e$, the optical depth $\tau$ 
is significantly lower in the one-T model than that of two-T models although
they have the same $\dot{M}$, because the one-T solution is of Bondi-type,
as we argued above, while the two-T solutions are disk-type, as shown by 
Fig. 6. If we could
find a disk-type one-T solution for $\dot{M}= \dot{M}_{\rm Edd}$,
the efficiency of this disk-type one-T model would be larger
than that of the two-T model. However, disk-type solutions for $\dot{M}=
\dot{M}_{\rm Edd}$ for one-T model are difficult to achieve because
the density and, thus, the radiative losses in this case would be so high
that the accretion flow would have collapsed at some radius, as shown by
the thin solid line for the one-T model in Fig. 4.

In summary, the properties of the one-T accretion models are
surprisingly similar to that of the two-T models, especially considering
that the value of parameter $\delta$ in two-T models is likely large,
$\delta\sim 0.5$ (Yuan, Quataert \& Narayan 2003). The temperature of one-T
model is only $\la 2$ times higher than that of a two-T model with
$\delta=0.5$. The maximum accretion rate of a one-T model below which a
global hot accretion solution can be obtained is only $\sim 3$ times
lower than that of the two-T model. A distinct difference between one
and two-T models is that the maximum luminosity of the former is $\sim
0.8\%L_{\rm Edd}$, while it is $\sim 16\%L_{\rm Edd}$ for a two-T
solution in our example for our given choice of $\alpha$.  Because the 
luminosity of the hard state of some BHCs is $>10\%L_{\rm Edd}$, this
makes the two-T model more likely for them (ref. Yuan \& Zdziarski
2004), unless much higher values of $\alpha$ are considered. 
An additional distinguishing characteristic between the two
types of models reflects the fact that the ADAF-LHAF transition for
one-T models occurs at $\dot{M}\sim 10^{-3}\alpha^2\dot{M}_{\rm Edd}$,
whereas it occurs at $\sim 10\alpha^2\dot{M}_{\rm Edd}$ for the two-T
models.  We note, however, that this distinguishing characteristic is
only theoretical since it has no effect on their emitted spectra as 
ADAF and LHAF solutions are smoothly connected. 

Given the similarity of the properties and dynamics between one and
two-T models, it is difficult to conclude purely on theoretical grounds
whether the one-T or two-T models are to be preferred. To determine their
viability, a comparison of the predicted spectra of the one-T model with
observations is warranted.  We choose two sources to carry out the
comparison; the supermassive black hole in the Galactic center, Sgr A*,
and the black hole binary, XTE J1118+480 in the following section. 
These sources are chosen because the observational data of these two 
sources are extensive and can provide constraints on the accretion models 
at both low and high luminosities (in units of Eddington luminosity) 
respectively. 

\section{Observational Tests}

\subsection{Sgr A*}

The spectral data of the quiescent state of Sgr A* over a range of
frequency from $10^9$ Hz to $10^{18}$ Hz is illustrated in Fig. 7.  The
description of the observational data as well as the theoretical fit to
this data using the most recent two-T ADAF model has been presented in
Yuan, Quataert \& Narayan (2003).  In this model, the accretion flow
starts from the Bondi radius, $R_{\rm B}$.  The accretion rate
($\dot{M}(R_{\rm B})\approx 10^{-5}\dot{M}_{\rm Edd}$) is calculated
from the Bondi formula using the observationally determined density and
temperature of the accretion flow at $R_{\rm B}$.  The dynamics of the
two-T ADAF has been calculated, taking into account the role of
outflow/convection ($s>0$ in eq. 1).  The emitted spectrum of the 
accretion flow has been calculated with the parameters adjusted to fit Sgr A*
as  presented in Yuan, Quataert \& Narayan (2003).  For our purpose, only
the synchrotron emission from the thermal electrons in the ADAF (with
edge-on viewing angle) is shown by the dashed line in Fig. 7. This is 
the only component responsible
for the high-frequency radio spectrum ($\ga 86$ GHz) in the model.
It can be seen from Fig. 7 that the theoretical spectrum
fits the high-frequency radio spectrum well. The low-frequency radio
below 86 GHz, which has a different slope with that above 86 GHz, 
infrared, and X-ray spectra, on the other hand, are 
reproduced by other components of the model involving nonthermal 
electrons in the ADAF (Yuan, Quataert \& Narayan 2003). To avoid 
introducing a nonthermal electron component, we focus on fitting the 
high frequency radio spectrum with the one-T accretion model. 

In fitting the high-frequency radio spectrum of Sgr A* to the results of a
one-T model, the same approach is adopted as in Yuan, Quataert \&
Narayan (2003). Because the synchrotron emission depends on the viewing
angle, we consider two extremes corresponding to face-on and edge-on.
The thick solid (face-on) and thin solid (edge-on) lines in Fig. 7 show the two
results for the one-T model. For the face-on viewing angle, in which 
the self absorption of the synchrotron emission is weaker, the predicted
radio flux is higher. The parameters are
$\dot{M}(R_{\rm B})\approx 10^{-5}\dot{M}_{\rm Edd}$, $\alpha=0.1,
\beta=0.9$\footnote{Note the difference of the definition of $\beta$
between Yuan, Quataert \& Narayan (2003) and the present paper.  There
$\beta$ was defined as the ratio between the gas pressure and the
magnetic pressure while here we define $\beta$ as the ratio between the
gas pressure to the total pressure (gas and magnetic).} and $s=0.45$. We
can see from the Fig. 7 that the thin solid line under-predicts the flux
of the radio peak while over-predicting the radio flux below the peak
significantly.  Increasing the value of $\dot{M}$ to fit the flux at the
peak leads to a poorer fit below the peak.  We point out that the 
results are insensitive to variations in the parameters such as $\alpha$, 
$\beta$ and $s$, indicating that our conclusion is robust.  The
physical reason for the over-prediction of the radio spectrum by the
one-T model is directly due to the fact that the temperature
in this model is higher than in the two-T model (ref. Fig. 1).

\subsection{XTE J1118+480}

To examine the goodness of fit of the theoretical spectrum with the observed 
spectrum at higher mass flow rates, the BH transient source 
XTE J1118+480 is considered.  Specifically, the archival RXTE 
data obtained during its 2000 outburst were analyzed. A total of 53 
observations were made of the source throughout the
outburst. We follow Cui (2004) closely in reducing the data,  constructing
X-ray spectra from the {\em Standard 2} data taken with the PCA detector,
as well as the corresponding background spectra from the latest background 
model for bright sources, for each of the detector units separately. Only 
data from the first xenon layer are used, in order to minimize calibration 
uncertainties, so the spectral coverage is limited roughly to 2.5--25 keV. 
The spectra are modeled in {\em XSPEC} (Arnaud 1996) and it is found that 
the spectra can be fitted satisfactorily with a power law, if a Gaussian 
function is included to model an apparent Fe $K_{\alpha}$ line. The 
best-fit models were used to derive the unfolded X-ray spectra. A set of 
representative spectra at different fluxes are shown in Fig.~8. For the 
highest-flux data set in the figure, the {\em HEXTE} data is also shown.
Moreover, a simultaneous {\em Chandra} observation was also re-analyzed
 to investigate a possible calibration artifact 
(as described in Yuan, Cui, \& Narayan 2005). We follow the standard CIAO 
threads (version 3.2) to derive the overall and background spectra from the 
LETG/ACIS observation, by making use of the latest calibration files 
(CALDB 3.0.1). For this work, only the first-order spectra are used and 
the positive and negative orders are co-added, excluding 
from the spectrum all known instrumental features as well as chip 
gaps, and modelled with a power law (with the hydrogen column density 
fixed at $N_H = 1.0 \times 10^{20}\mbox{ }{\rm cm}^{-2}$) in {\em XSPEC}. 
The simple model provides an adequate (but not statistically acceptable) 
fit to the data. The most notable feature in the residuals is a sharp
drop at $\sim$2 keV, which we assume is caused by calibration uncertainty.
We exclude the feature and re-fit the spectrum. The unfolded (and 
unabsorbed) spectrum is also shown in Fig.~8, along with the measurements 
in other wavebands. It is reassuring that the previously seen spectral 
break (McClintock et al. 2001) is no longer present. We also note that 
there is a general agreement between the independently derived spectral 
results for the {\em Chandra} and {\em RXTE} observations.

Compared to Sgr A*, the mass of the
black hole in XTE J1118+480 is 5 orders of magnitude lower and for the
highest flux level of XTE J1118+480, the bolometric luminosity (in units
of Eddington rate) is about 4 orders of magnitude higher.  In this case,
the density of the accretion flow is $\sim 10$ orders of magnitude 
higher and the thermal coupling between ions and electrons may be expected 
to be more effective. As these physical conditions differ significantly 
from that of Sgr A*, it is necessary to check whether the one-T model is 
viable in this mass flow rate regime. 

In contrast to the case for Sgr A*, we focus on the X-ray spectrum of 
XTE J1118+480 since this part of the spectrum is not affected by 
other components in the system.  That is, the X-ray spectrum 
is produced from the emission from the hot accretion flow with negligible
contributions from the standard thin disk and a hypothesized jet 
(see Yuan, Cui \& Narayan 2005 and Yuan \& Cui 2005 for details). 
The solid lines in Fig. 8 show the results of the fitting of the one-T
models to the observed spectra assuming that the same comparison can 
apply for these models. From the top to the bottom in Fig. 8,
the parameters are: $\alpha=0.3$, $\beta=0.9$, $\dot{M}=0.6, 0.2,
0.1$, and $0.04$ $\dot{M}_{\rm Edd}$.  Following the assumption in
Yuan, Cui \& Narayan (2005), we assume that there is no outflow when the
advection factor is negative, i.e., $s=0$ so that the mass flow rate is
constant in this spatial region.  It can be seen that the one-T model
fails to fit the X-ray data at all the four flux levels. Specifically, 
the X-ray spectrum predicted by the one-T model exhibits exaggerated bumps 
in the spectrum in comparison to the observed spectra at the lower flux levels.
As a comparison, we show in Fig. 8 the results of fitting the two-T model by
the four dashed lines (see Yuan, Cui \& Narayan 2005 for the detail of the
model and its difference from Esin et al. 2001).
The parameters are $\alpha=0.3, \beta=0.9,
\delta=0.5$, $\dot{M}=0.5, 0.15, 0.09,$ and $0.01 \dot{M}_{\rm Edd}$.
We see that they can reasonably fit the observational 
data\footnote{One {\em BeppoSAX} observation (Frontera et al. 2001)
show a high-energy break at $\ga$ 100 keV at the highest-flux data set. 
Our two-T model predicts a cut-off at $\sim 200$ keV, which is somewhat larger
than the above value. The disprepancy may be partly due to the fact 
that the {\em BeppoSAX} observation
was not obtained simultaneously with other telescopes. In fact, there seems to 
be no clear evidence of cutoff below 200 keV in the later two {\em BeppoSAX} 
observations of the same source at similar flux level (ref. Fig. 3 in 
Frontera et al. 2003). On the other
hand, it is interesting to investigate whether the two-T model can 
produce a low energy cutoff in this source.} 
The physical reason for the discrepancy of the fit
between the one-T and two-T models is similar to the discrepancy for Sgr
A*, namely, the one-T model predicts a higher temperature (ref. Figs.  2
\& 3) than the corresponding two-T model. Therefore, the Compton 
scattering ``bump'' in the spectrum predicted by the one-T model is more 
obvious than that produced by the two-T model at the lower flux levels. 
Thus, taking into account both the observations of XTE J1118+480 and Sgr A*, 
we find that the hot one-T model is not likely applicable for explaining 
the observed spectra. 

\section{Summary}

The majority of hot accretion flow models (such as ADAFs) are based on
the assumption that the only coupling mechanism between the ions and
electrons involves Coulomb collisions. Since it is ineffective for the
physical conditions in these flows, the ions are much hotter than the
electrons, i.e., the hot accretion flow is two-T.  However, the hot
accretion flow is, in general, collisionless and collective plasma
effects are likely to be significant.  Hence, the ions and electrons may
be strongly thermally coupled, having a common temperature. In the
present paper, we have investigated the hot accretion flows in the one-T
approximation and determined its viability by comparison of theoretical
spectra with those observed. 

It is found that the dynamics of the one-T hot accretion flow are quite
similar to that of the two-T models for large fractions, $\delta$
($\delta \sim 0.5$), of the viscous energy directly heating the
electrons. Specifically, the temperatures predicted by the one-T model
are only $\sim 2$ times higher than that of the two-T model with the
same parameters (ref. Figs 1-4).  The maximum accretion rate below which
the hot one-T accretion solutions exist confirms the $\alpha^2$ scaling 
of previous authors, but is much higher, of the order of $10\sim \alpha^2 
\dot{M}_{\rm Edd}$, which is only $\sim 3$ times lower than that for 
two-T models (Fig. 4).  Furthermore, the radiative efficiency of the one-T model   
can be comparable to that of two-T models (Fig. 5). On the other hand, the
electron temperature and radiative efficiency of the two-T model with
$\delta=0.01$ is much smaller than than of the one-T model over the same
mass flow rate range. The main dynamical difference between one and
two-T models stems from the fact that the advection factor (defined in
eq. (8)) becomes negative (corresponding to advective heating) when
$\dot{M}$ is higher than $\sim 10^{-3}\alpha^2\dot{M}_{\rm Edd}$ for the
one-T models, while this transition occurs at $\sim 10\alpha^2\dot{M}_{\rm
Edd}$ for the two-T models.

In order to determine the effectiveness of the thermal coupling between
the ions and electrons, we have adopted a phenomenological approach and
attempted to fit the observed spectra of the black hole sources, Sgr A*
and XTE J1118+480 using the one-T model. It has been shown in previous
works that two-T models can fit the spectra of these two sources (and
many other features; see Yuan, Quataert \& Narayan 2003 for Sgr A* and
Yuan, Cui \& Narayan 2005 for XTE J1118+480) adequately.  Our
comparisons of observed spectra with theoretical spectra of the one-T
model reveal that the fits to Sgr A* in the radio regime and 
XTE J1118+480 in the X-ray regime are not satisfactory (Figs. 7, 8).  
Although the spectral fitting technique is not without assumptions, 
our results appear to be robust and indicate that this inadequacy is a 
direct result of the high temperatures in the one-T model. Thus, black 
hole systems are likely described by a hot accretion flow in a two-T 
configuration, suggesting 
that the thermal coupling between ions and electrons is not complete.  

The large fraction of turbulent
viscous energy required for direct heating to the electrons remains to
be understood.  Although the physical mechanism has been investigated
(e.g., Quatatert \& Gruzinov 1999), theoretical
uncertainties still exist.  As a result, determinations of $\delta$
will continue to require a phenomenological approach.

\begin{acknowledgements}
We thank the anonymous referee for his/her very constructive suggestions.
This work was supported in part by NASA grants NAG5-9998, NNG04GI54G, 
NNG05GF91G and the One-Hundred-Talent Program of China.
\end{acknowledgements}


\clearpage

\begin{figure} \epsscale{1.0} \plotone{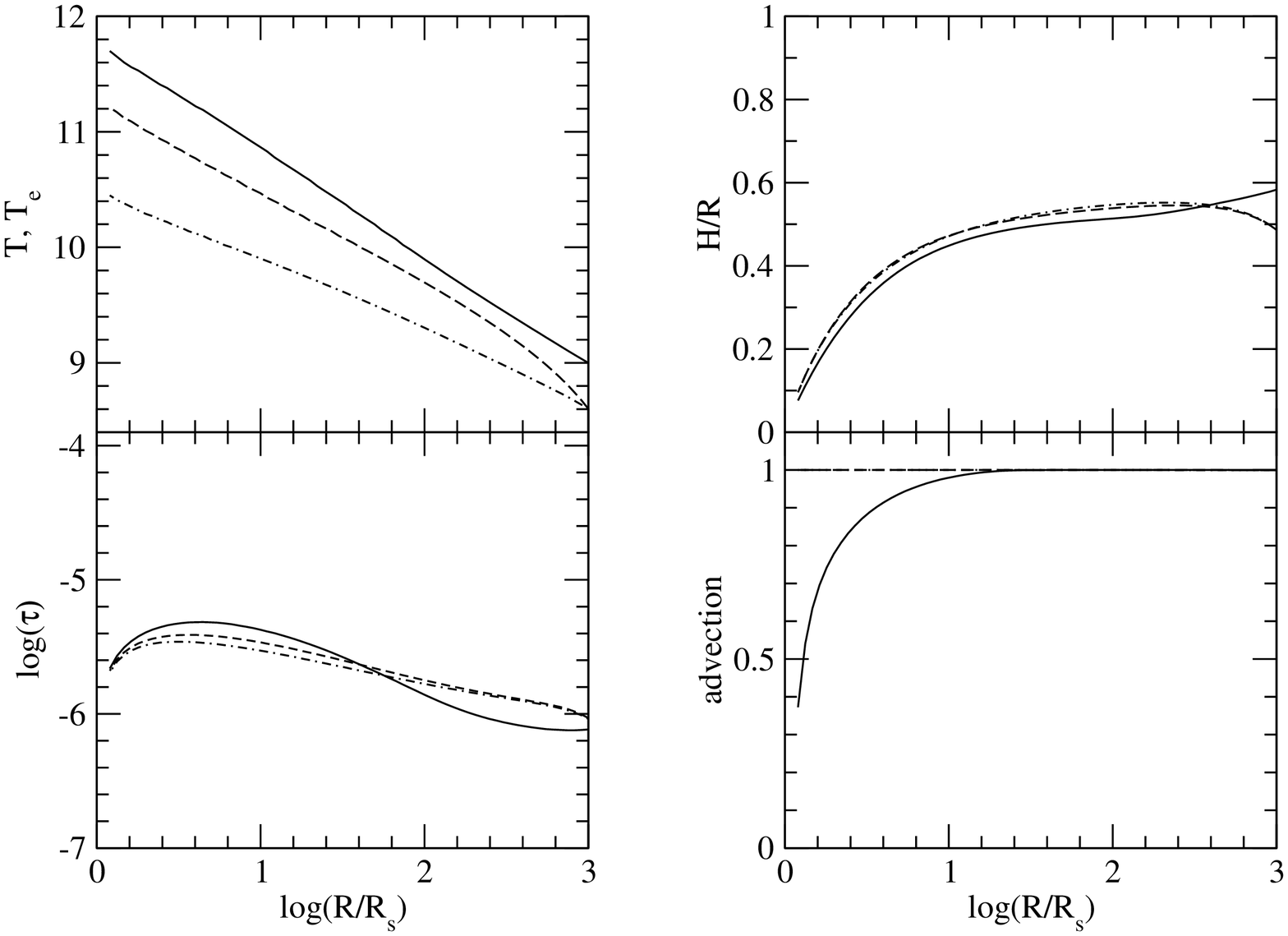} \vspace{.0in}
\caption{The profiles of temperature, optical depth, ratio of scale
height to radius, and the advection factor of a hot one-temperature
accretion solution (solid lines).  The parameters are: $M=10\msun,
\dot{M}=10^{-5} \dot{M}_{\rm Edd}, \alpha=0.3, \beta=0.9$. The outer
boundary conditions are $R_{\rm out}=10^3R_s, T=10^9 K, v/c_s=0.5$. For
simplicity, effects associated with outflow or convection are not taken
into account. The two-temperature solutions with the same parameters and
$\delta =0.5$ (dashed lines) and $0.01$ (dot-dashed lines) are also
shown for comparison.} \end{figure} 

\clearpage \begin{figure}
\epsscale{1.0} \plotone{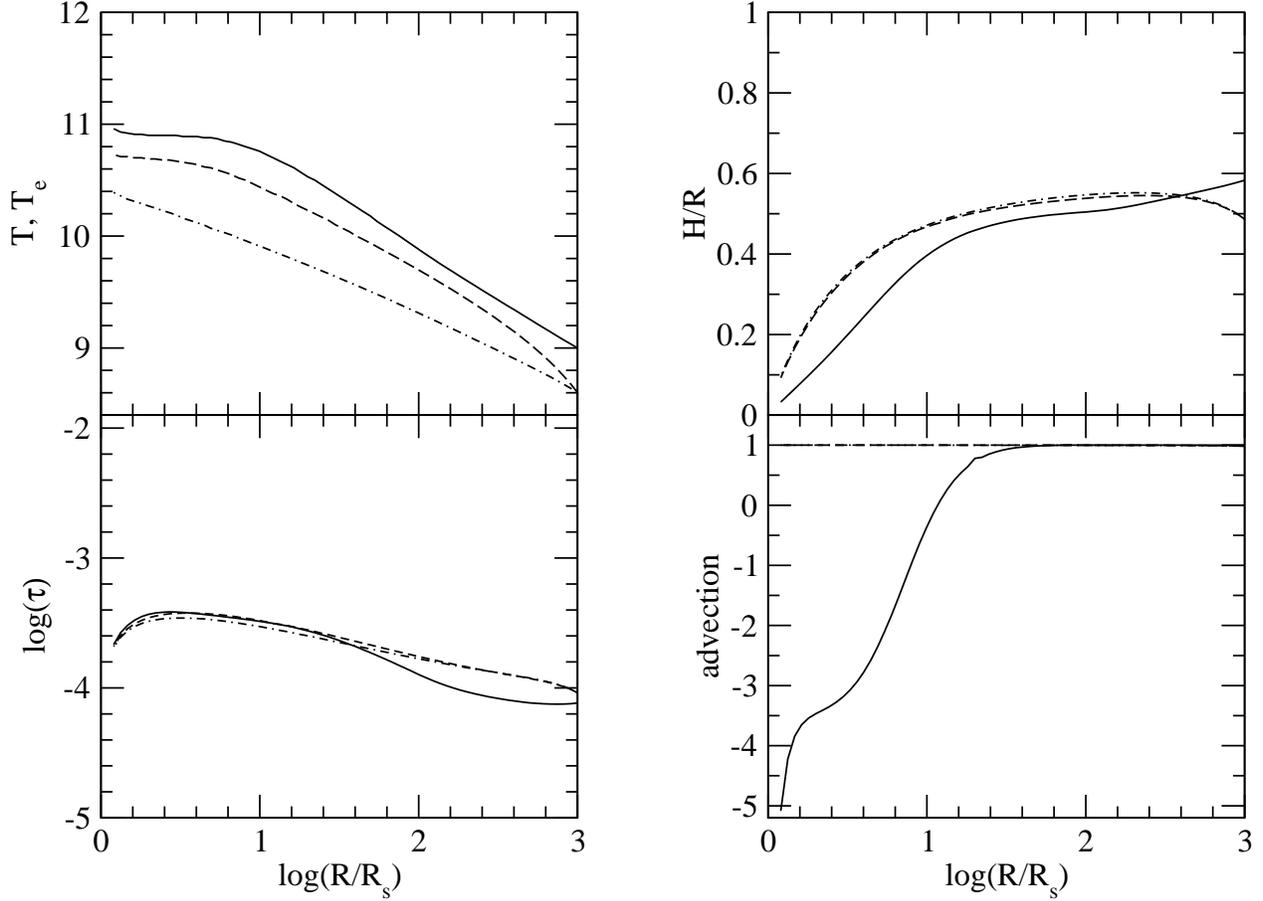} \vspace{.3in} \caption{Same as Fig.
1, except for $\dot{M}=10^{-3}\dot{M}_{\rm Edd}$.  Note that the
advection factor of the one-temperature solution at the inner region
becomes negative implying that advective heating is important in this
region. Hence, the inner region of the solution belongs to an LHAF
rather than an ADAF. On the other hand, the two-temperature solution
remains an ADAF.} \end{figure} 

\clearpage \begin{figure} \epsscale{1.0}
\plotone{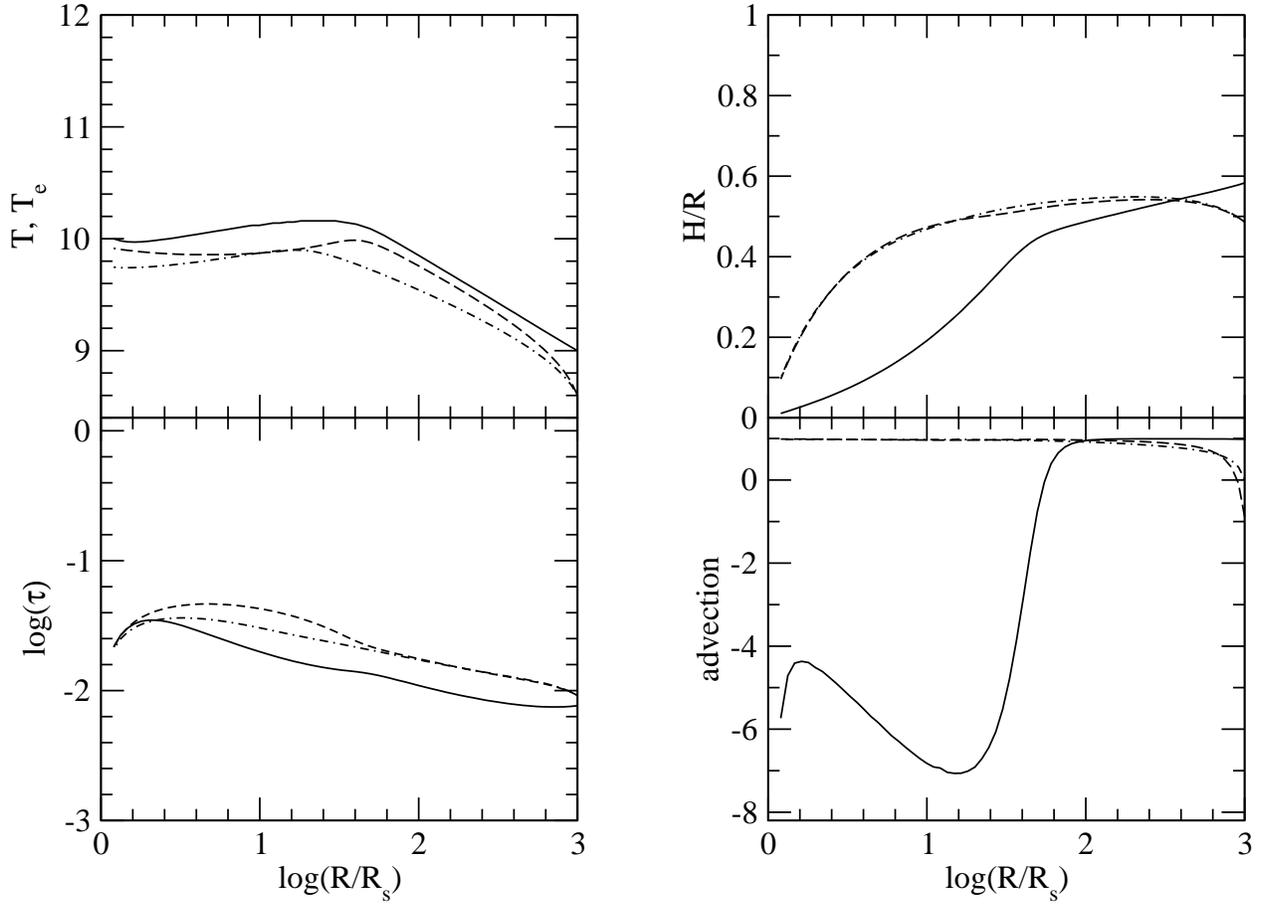} \vspace{.2in} \caption{Same as Fig. 1, except for
$\dot{M}=10^{-1}\dot{M}_{\rm Edd}$.  Note that the LHAF region increases
in spatial extent due to the stronger radiative cooling.} \end{figure}

\clearpage \begin{figure} \epsscale{0.90} \plotone{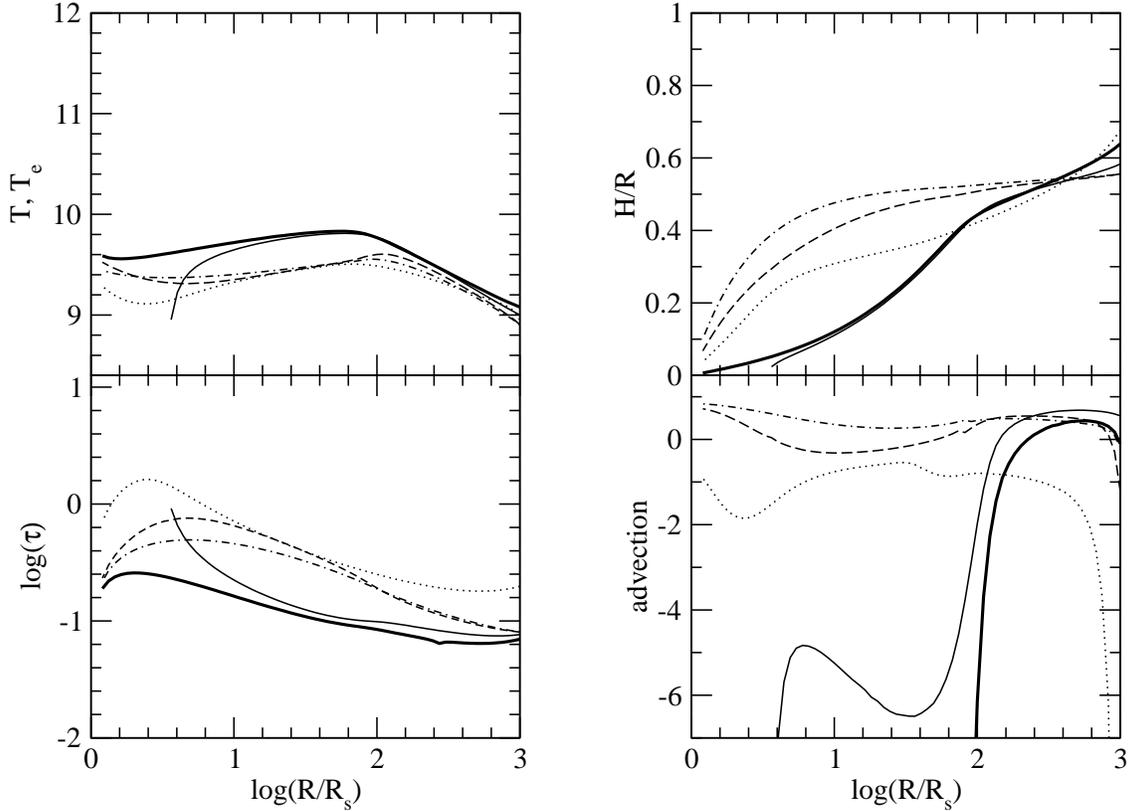}
\vspace{.2in} \caption{Similar to Figs. 1-3. The thick and thin solid
lines are for one-temperature Bondi-type and disk-type solutions
respectively for $\dot{M}=\dot{M}_{\rm Edd}$ corresponding to
different outer boundary conditions.  This value of $\dot{M}$ is roughly
the maximum rate below which the hot one-temperature solution exists. 
Above this value, the inner region of the flow collapses because of the
strong cooling (similar to the solution described by the thin solid
line). The dashed and dot-dashed lines are for two-temperature solutions
with $\dot{M}=\dot{M}_{\rm Edd}$ for $\delta = 0.5$ and 0.01
respectively.  This rate corresponds to the critical rate of ADAF above
which it becomes LHAF. The dotted line shows an example of a
two-temperature LHAF solution with $\dot{M}=3\dot{M}_{\rm Edd}$.}

\end{figure}

\clearpage 
\begin{figure} 
\epsscale{1.0} 
\plotone{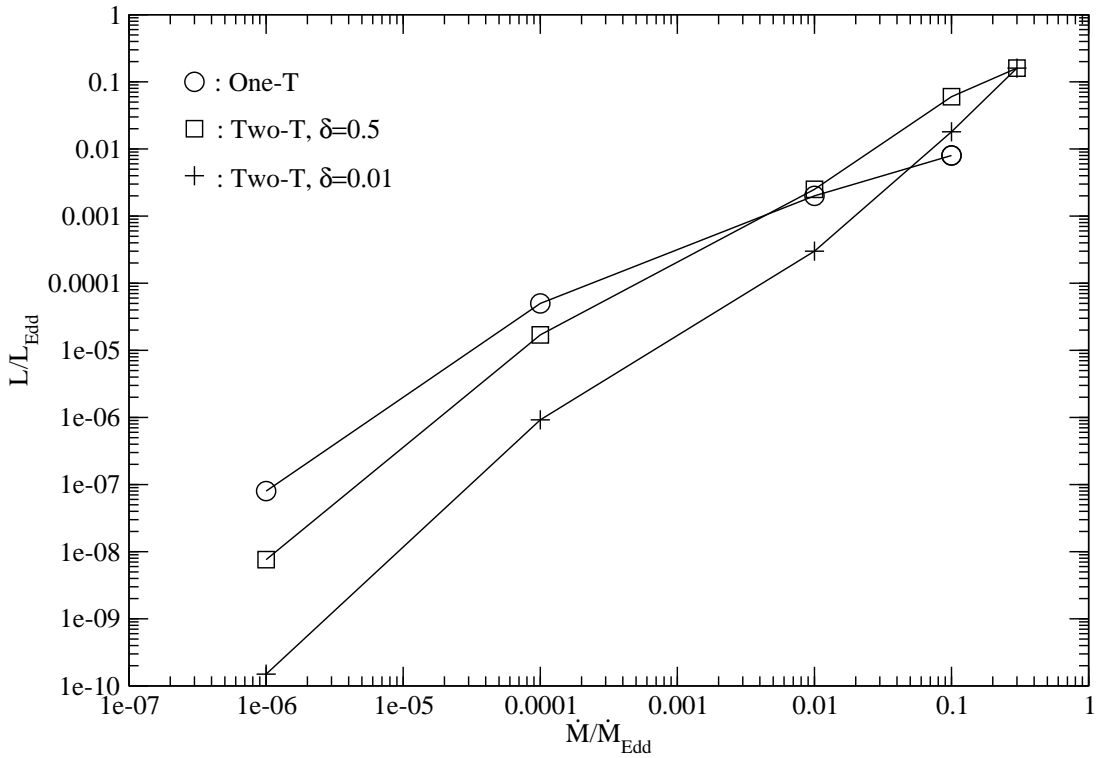} 
\vspace{.2in}
\caption{The emitted bolometric luminosity relative to the Eddington
luminosity as a function of the mass flow rate normalized to the
Eddington value for the solutions presented in Figs. 1-4. The data 
point for the one-T solution with $\dot{M}/\dot{M}_{\rm Edd}=1$
is only for the thick solid line in Fig. 4.} 
\end{figure}

\clearpage 
\begin{figure} 
\epsscale{1.0} 
\plotone{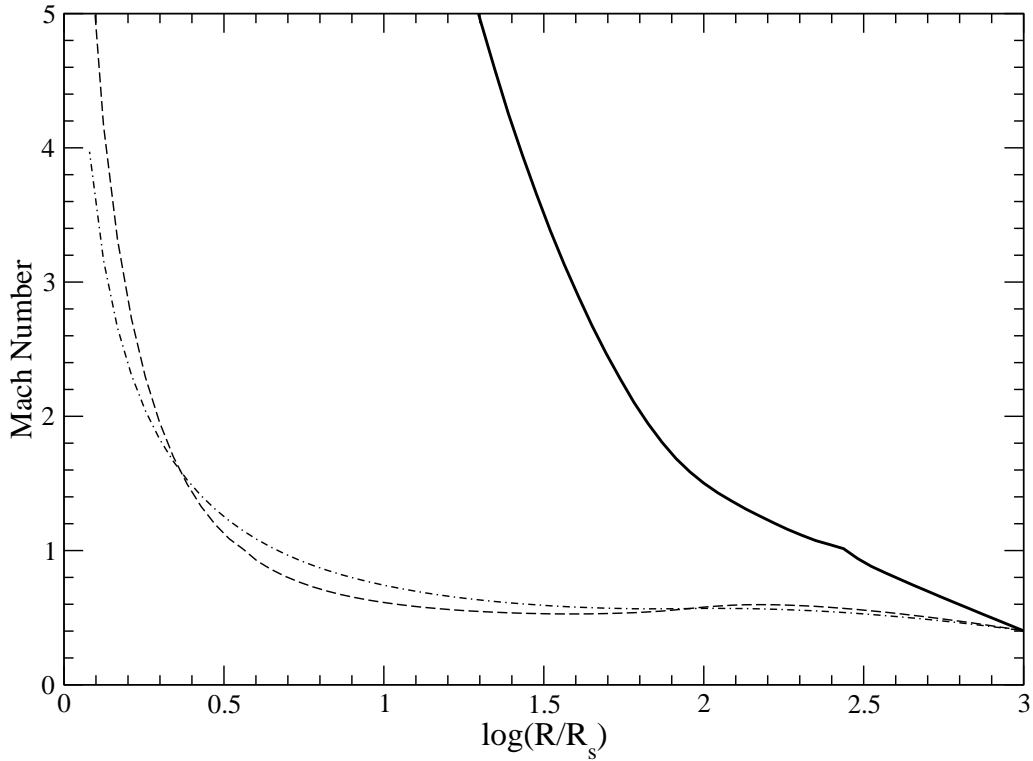}
\vspace{.2in} 
\caption{The variation of the Mach number (the ratio of
the radial drift velocity to the speed of sound) as a function of radius
(normalized to the Schwarzschild radius) for the corresponding three
solutions presented in Fig. 4. The solution denoted by the thick solid
line is of a Bondi-type with a large sonic radius while the other two
are disk-type.  These different flow descriptions provide an
understanding of the low optical depth corresponding to the thick
solid line in comparison to the high densities denoted by the dashed and
dot-dashed lines for the same value of $\dot{M}$.} 
\end{figure}

\clearpage
\begin{figure}
\epsscale{1.0}
\plotone{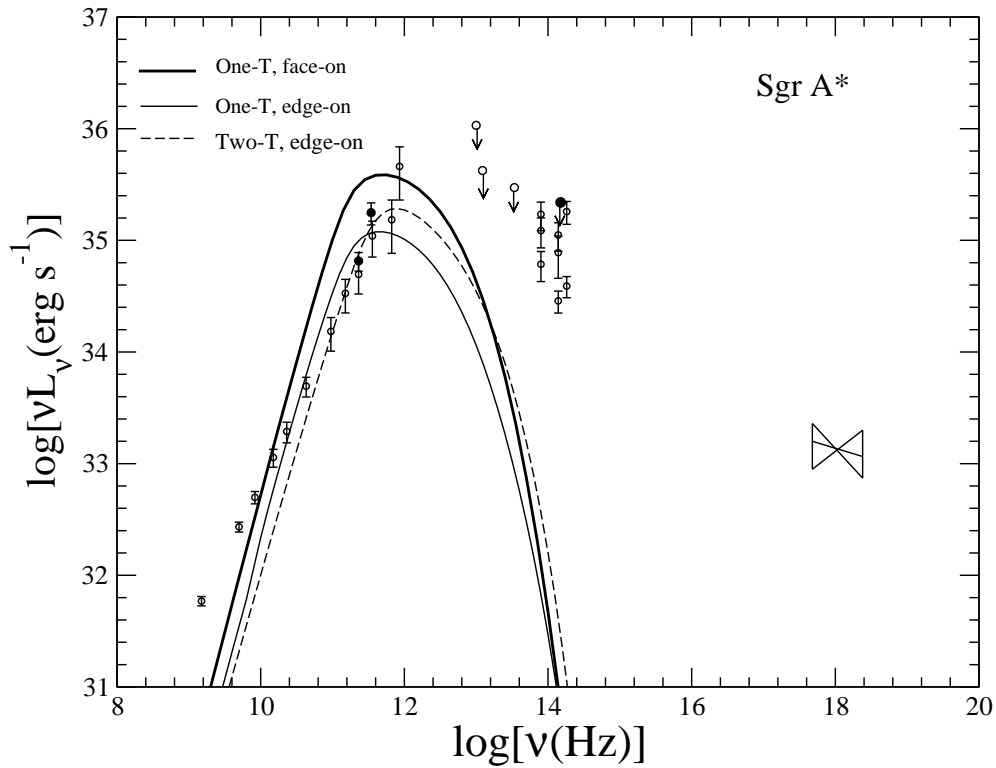}
\vspace{.2in}
\caption{The modeling results for one and two-temperature hot accretion
flows for the quiescent state of the supermassive black hole Sgr A*. Two
results for the one-temperature solution are shown, corresponding to
face on (thin solid line) and edge on (dashed line) viewing angles.  It
can be seen that the one-temperature models over-predicts the flux level
in the high-frequency radio (above $\sim 86$ GHz), while the two-temperature 
solution fits the spectrum.}
\end{figure}

\clearpage
\begin{figure}
\epsscale{0.9}
\plotone{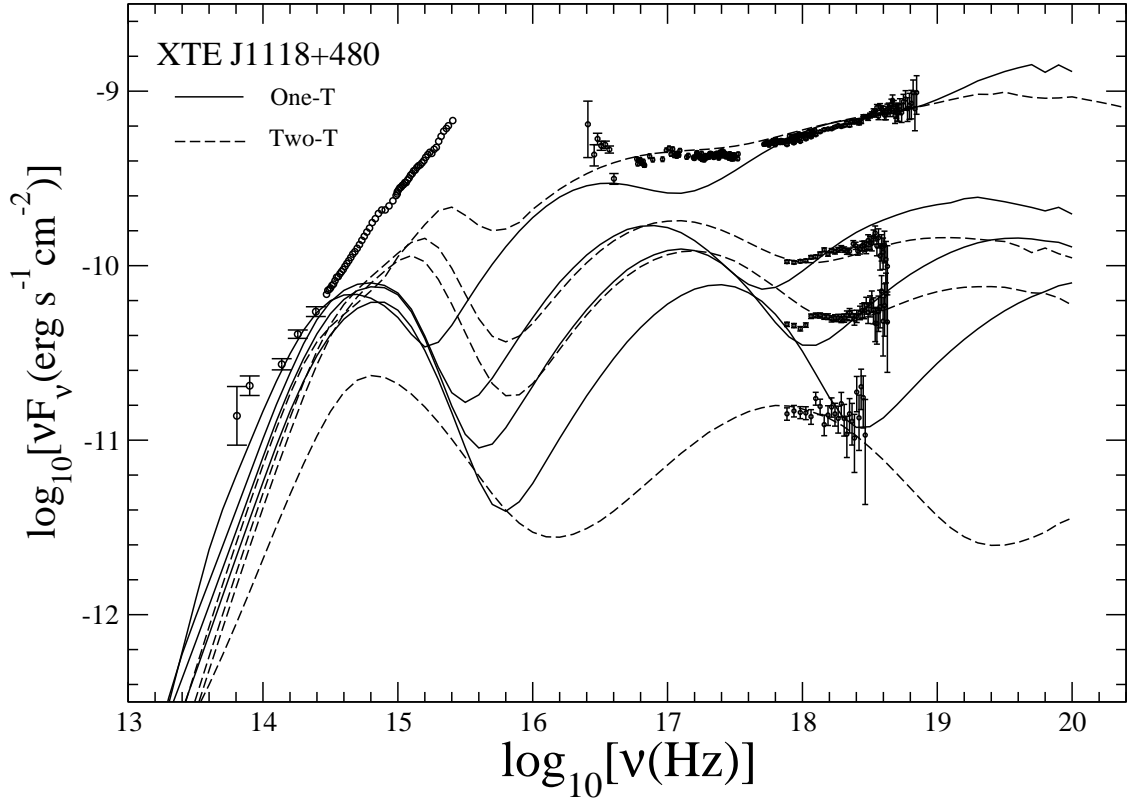}
\vspace{.2in}
\caption{The modeling results for one (solid line) and two-temperature
(dashed line) hot accretion flows to the black hole candidate, XTE
J1118+480, in its 2000 outburst at different flux levels. The emissions 
from the thin disk and the jet are not included since their contributions
to the X-ray spectrum are negligible, as shown by Yuan, Cui \& Narayan (2005). 
We can see that the two-temperature models can fit the spectra well, while the
one-temperature does not, because the predicted electron temperatures
are too high.}
\end{figure} 

\end{document}